# Drugst.One - A plug-and-play solution for online systems medicine and network-based drug repurposing

Andreas Maier[1,*], Michael Hartung[1,*], Mark Abovsky[2,3], Klaudia Adamowicz[1], Gary D. Bader[4,5,6,7,8], Sylvie Baier[9], David B. Blumenthal[10], Jing Chen[11], Maria L. Elkjaer[12,13,14], Carlos Garcia-Hernandez[15], Mohamed Helmy[4,5,6], Markus Hoffmann[9,16,17], Igor Jurisica[2,3,18,19], Max Kotlyar[2,3], Olga Lazareva[20,21,22], Hagai Levi[23], Markus List[9], Sebastian Lobentanzer[24], Joseph Loscalzo[25], Noel Malod-Dognin[15], Quirin Manz[9], Julian Matschinske[1,9], Miles Mee[4,5,6], Mhaned Oubounyt[1], Alexander R. Pico[26], Rudolf T. Pillich[11], Julian M. Poschenrieder[1,9], Dexter Pratt[11], Nataša Pržulj[15,27,28], Sepideh Sadegh[1,9,29], Julio Saez-Rodriguez[24], Suryadipto Sarkar[10], Gideon Shaked[23], Ron Shamir[23], Nico Trummer[9], Ugur Turhan[1], Ruisheng Wang[25], Olga Zolotareva[1,9], Jan Baumbach[1,30,†]

[1] Institute for Computational Systems Biology, University of Hamburg, Hamburg, Germany
[2] Division of Orthopaedic Surgery, Schroeder Arthritis Institute, and Data Science Discovery Centre, Osteoarthritis Research Program, Krembil Research Institute, UHN, Toronto, Canada
[3] Data Science Discovery Centre for Chronic Diseases, Krembil Research Institute, University Health Network, 60 Leonard Avenue, 5KD-407, Toronto, ON, M5T 0S8, Canada
[4] Department of Molecular Genetics, University of Toronto, Toronto, ON, Canada
[5] The Donnelly Centre, University of Toronto, Toronto, ON, Canada
[6] Department of Computer Science, University of Toronto, Toronto, ON, Canada
[7] Princess Margaret Cancer Centre, University Health Network, Toronto, ON, Canada
[8] The Lunenfeld-Tanenbaum Research Institute, Mount Sinai Hospital, Toronto, ON, Canada
[9] Chair of Experimental Bioinformatics, TUM School of Life Sciences, Technical University of Munich, Munich, Germany
[10] Department Artificial Intelligence in Biomedical Engineering (AIBE), Friedrich-Alexander University Erlangen-Nürnberg (FAU), 91052 Erlangen, Germany
[11] Department of Medicine, University of California San Diego, 9500 Gilman Drive, La Jolla, CA, 92093, USA
[12] Department of Neurology, Odense University Hospital, Odense, Denmark
[13] Institute of Clinical Research, University of Southern Denmark, Odense, Denmark
[14] Institute of Molecular Medicine, University of Southern Denmark, Odense, Denmark
[15] Barcelona Supercomputing Center (BSC), 08034 Barcelona, Spain
[16] Institute for Advanced Study (Lichtenbergstrasse 2a, D-85748 Garching, Germany), Technical University of Munich, Germany
[17] National Institute of Diabetes, Digestive, and Kidney Diseases, Bethesda, MD 20892, United States of America
[18] Departments of Medical Biophysics and Computer Science, University of Toronto, Toronto, Canada
[19] Institute of Neuroimmunology, Slovak Academy of Sciences, Bratislava, Slovakia
[20] Division of Computational Genomics and Systems Genetics, German Cancer Research Center (DKFZ), 69120 Heidelberg, Germany
[21] Junior Clinical Cooperation Unit Multiparametric methods for early detection of prostate cancer, German Cancer Research Center (DKFZ), Heidelberg, Germany
[22] European Molecular Biology Laboratory, Genome Biology Unit, 69117 Heidelberg, Germany
[23] Blavatnik School of Computer Science, Tel-Aviv University, Tel-Aviv, Israel
[24] Heidelberg University, Faculty of Medicine, and Heidelberg University Hospital, Institute for Computational Biomedicine, Bioquant, Heidelberg, Germany

[25] Department of Medicine, Brigham and Women's Hospital, Harvard Medical School, Boston, MA 02115, USA
[26] Institute of Data Science and Biotechnology, Gladstone Institutes, 1650 Owens Street, San Francisco, 94158, California, USA
[27] Department of Computer Science, University College London, London WC1E 6BT, UK
[28] ICREA, Pg. Lluís Companys 23, 08010 Barcelona, Spain
[29] Department of Clinical Genetics, Odense University Hospital, Odense, Denmark
[30] Computational Biomedicine Lab, Department of Mathematics and Computer Science, University of Southern Denmark, Odense, Denmark

[*] These authors have contributed equally
[†] These authors supervised the project

(Authors except first and last are sorted alphabetically by last name)

## Abstract

In recent decades, the development of new drugs has become increasingly expensive and inefficient, and the molecular mechanisms of most pharmaceuticals remain poorly understood. In response, computational systems and network medicine tools have emerged to identify potential drug repurposing candidates. However, these tools often require complex installation and lack intuitive visual network mining capabilities. To tackle these challenges, we introduce Drugst.One, a platform that assists specialized computational medicine tools in becoming user-friendly, web-based utilities for drug repurposing. With just three lines of code, Drugst.One turns any systems biology software into an interactive web tool for modeling and analyzing complex protein-drug-disease networks. Demonstrating its broad adaptability, Drugst.One has been successfully integrated with 21 computational systems medicine tools. Available at https://drugst.one, Drugst.One has significant potential for streamlining the drug discovery process, allowing researchers to focus on essential aspects of pharmaceutical treatment research.

## Introduction

In recent years, rapid technological advancements and unmet medical needs have fueled the development of computational tools that leverage systems biology methodologies to decipher complex biomedical data [1]. These tools frequently target the identification of specific proteins or genes in a given disease context, such as marker genes indicative of disease progression [2,3]. The visualization of these results in a biomedical network context can greatly improve their interpretability, allowing us to better understand the underlying disease mechanisms and the interrelationships among the identified entities [4]. This principle applies to a variety of biomedical fields, including oncology [5,6], virology [7], and disease subtype identification and patient stratification through differential gene expression analysis [8,9]. Rendering these intricate cellular processes as graphs aids researchers in tailoring more precise pharmaceutical treatments, minimizing side effects [10,11] and opening prospects for novel therapeutic and diagnostic strategies, such as mechanistic drug repurposing [12].

Key challenges in the development of systems biology platforms include the integration of comprehensive biomedical data and the creation of flexible graphical user interfaces for data analysis, prioritization, and visualization. Stand-alone software such as Cytoscape [13] visualizes biological networks but necessitates local installation for each user. To circumvent this, developers often provide online solutions dependent solely on browser compatibility. However, this presents additional hurdles for researchers who may lack sufficient web-development skills and need to establish and maintain an infrastructure, including a server hosting a database and a website. Beyond network visualization, the collection, harmonization, integration, and incorporation of diverse biomedical data demand a significant time investment [14]. Moreover, the database should be maintained and regularly updated, a chore that is often not addressed by bioinformatics tools that primarily provide a result overview with a limited set of features. Thus, if network exploration is not neglected due to the additional workload, unique solutions are being developed from scratch, resulting in network visualizers and explorers of varying quality [7,8,15].

We developed Drugst.One to reduce software engineering overhead, bundle development capacities, and to standardize and simplify network analysis and visual network exploration for biomedical web tools (Figure 1). With minimal programming effort, Drugst.One can turn any gene or protein-based systems biology tool into a powerful online toolkit for network integration and visualization, as well as mechanistic drug repurposing. Drugst.One is a customizable plug-and-play solution for web-application developers in need of a feature-rich network explorer coupled with a biomedical protein-drug-disease network data warehouse. With as little as three lines of code, Drugst.One can be added to any biomedical web tool, highlighting opportunities for drug repurposing and elucidating disease mechanisms. Incorporating multiple state-of-the-art databases (see Supplementary Table 1) to complement visualized data, Drugst.One provides an intuitive interface for applying algorithms for exploratory network analyses, drug target and drug repurposing candidate identification and prioritization. Weekly updates guarantee the relevance of its database for frequently changing data. Currently, Drugst.One is integrated into 21 systems medicine software resources (Table 1), including mirDIP (see Supplementary Figure S3) [16] and WikiPathways [17]. In this article, we describe the functionality of Drugst.One and demonstrate its utility on the basis of two studies – on drug repurposing for inflammatory bowel disease (IBD) and on exploring the smooth muscle cell (SMC) proliferation pathway.

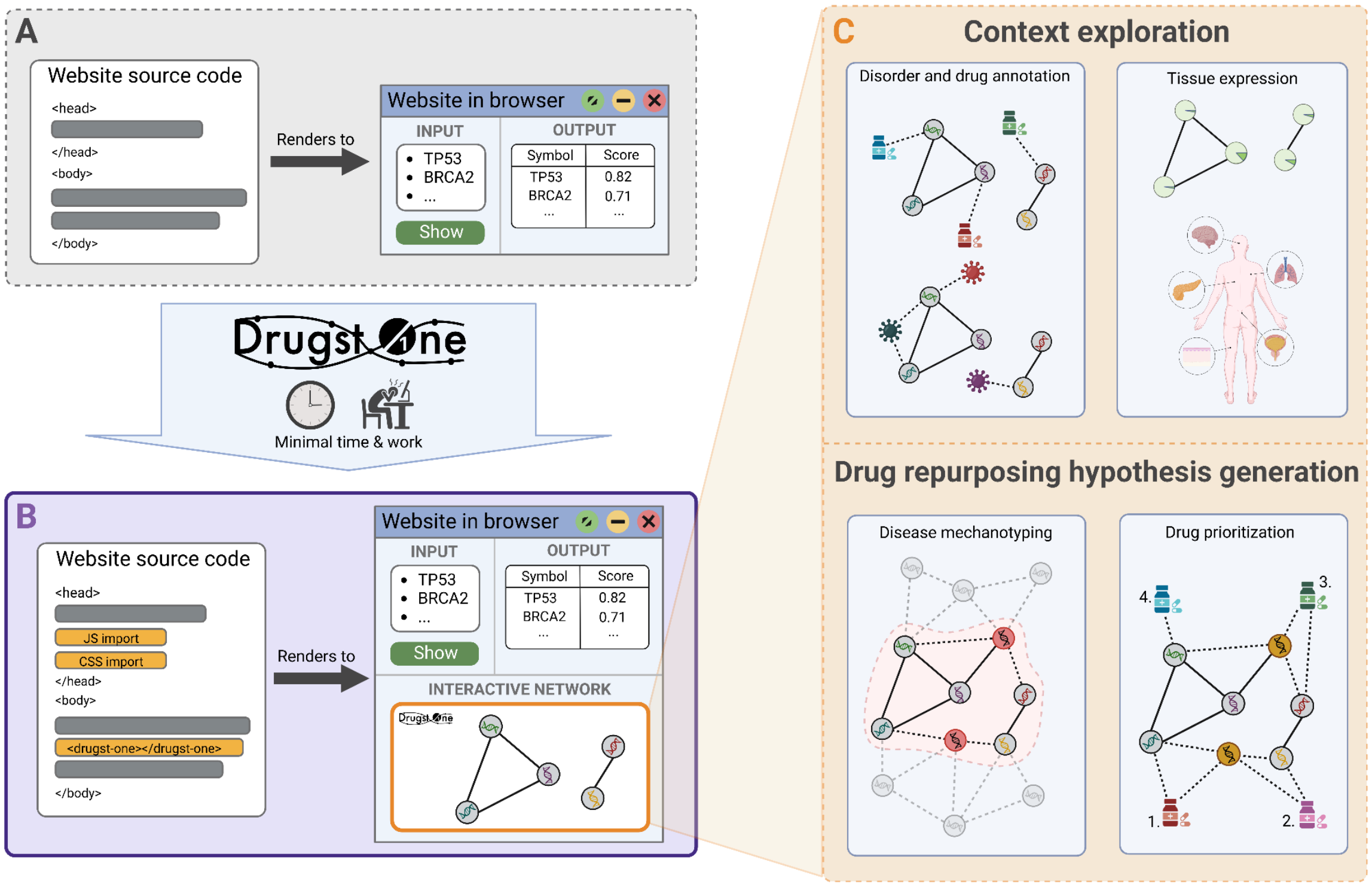

**Figure 1:** Drugst.One enables web developers to add a fully functional network explorer to any website with minimal coding effort (biomedical web tool before (**A**) and after (**B**) Drugst.One integration). (**C**)**:** A network can be explored manually or by using network medicine algorithms to identify disease mechanisms and drug repurposing candidates. Associated diseases and tissue-specific expression are additional information layers to gain insight into the network context.

## Results

### Drugst.One overview

Drugst.One closes the gap between disease mechanism mining and hypothesis generation for drug repurposing. The required input is a list of proteins or genes in HGNC, UniProt, Ensembl, or Entrez ID space. On demand, these entities are integrated into the interactome and automatically annotated with clinically relevant information, e.g., targeting drugs or known disease associations. Exploratory functions allow the visualization of known drug indications and disease associations as well as an overlay for tissue-specific expression information (Figure 1C). For most information-enriching functions, Drugst.One provides several data sources to choose from (Supplementary Notes 3.1). Convenience features for network control (such as enabling the interactive mode or resetting the view) and export are available to assist exploratory analysis further.

Drugst.One originates from the network-based drug repurposing platforms CoVex [7] and CADDIE [5], developed for the application in SARS-CoV-2 and cancer, respectively. While they provide disease-specific information, both tools share underlying principles and algorithms. These tested and published methods form the Drugst.One algorithmic toolkit for more extensive analysis. Module identification algorithms provide means to identify additional potential drug targets from the interactome to enrich the mechanistic context. In a second step, drugs that are directly or indirectly linked can be ranked. This allows the assessment of the compound's potential to be repurposed using network-based algorithms. Although both steps work automatically, users can infuse their expert knowledge by adjusting input gene sets. Users can choose among seven drug prioritization and drug target identification algorithms to rank small molecules directly or indirectly targeting disease proteins, thus serving as potential drug repurposing candidates (Supplementary Notes 3.2).

Overrepresentation analysis using g:Profiler [18] or functional coherence validation using DIGEST [19] on all loaded proteins in the network can be run with one click. Further, searching for curated pathways containing the same proteins in NDEx IQuery [20] allows for even more interoperability. A full list of projects partnering with the ‘Drugst.One Initiative’ by integrating Drugst.One can be found in Table 1. Collaborators that assisted and provided technologies that helped to build features in Drugst.One can be found in Supplementary Notes 4.

| Tool | URL | Tool Description | Integration | Integration status |
|---|---|---|---|---|
| BiCoN [8] | https://exbio.wzw.tum.de/bicon/ | Network-constrained patient stratification through biclustering | Plugin | Done |
| DOMINO [21] | http://domino.cs.tau.ac.il/ | Active module identification with improved empirical validation | Link-out | In progress |
| G-Browser | https://exbio. | An enhanced genome | Plugin | Done |

| Tool | URL | Tool Description | Integration | Integration status |
|---|---|---|---|---|
| | wzw.tum.de/genome-browser/ | browser plugin that seamlessly integrates data sources and functions for genetics research. | | |
| GraphFusion | https://github.com/CarlosJesusGH/GraphFusion | An intuitive web-based graph analytics, fusion, and visualization tool | Plugin | Done |
| GraphSimViz [22] | https://graphsimviz.net/ | Visualization of diseasomes, drugomes, and drug-disease networks | Plugin | Done |
| HitSeekR [23] | https://exbio.wzw.tum.de/hitseekr/ | User-friendly tool for drug (target) identification in high-throughput screening | Plugin | Done |
| Interactive Enrichment Analysis [24] | https://github.com/gladstone-institutes/Interactive-Enrichment-Analysis/ | Enrichment analysis on multiple public datasets | Link-out | In progress |
| mirDIP [16] | https://ophid.utoronto.ca/mirDIP/ | Integrated microRNA-target data integration portal | Plugin | Done |
| NAViGaTOR [25] | https://ophid.utoronto.ca/navigator/ | Network visualization and analysis software | Link-out | In progress |
| NDEx IQuery [20] | https://www.ndexbio.org/iquery/ | Web tool for pathway and network-based gene set analysis | Plugin | Planned |
| NeEDL - Epistasis Disease Atlas | https://epistasis-disease-atlas.com | Web resource to visualize, investigate, and interpret higher-order genetic interactions of single nucleotide polymorphisms in 18 human heritable diseases. | Link-out | Done |
| NeEDL - R Shiny App | https://hub.docker.com/r/bigdatainbiomedicine/needl | R shiny app to visualize, investigate, and interpret higher-order genetic interactions of single nucleotide polymorphisms on locally computed datasets. | Plugin | Done |

| Tool | URL | Tool Description | Integration | Integration status |
|---|---|---|---|---|
| openPIP [26] | https://github.com/BaderLab/openPIP | Open platform to store and retrieve protein-protein interaction datasets. | Link-out | Done |
| pathDIP [27] | https://ophid.utoronto.ca/pathDIP | Integrated pathway database and pathway enrichment analysis portal | Plugin | In progress |
| Pathway Figure OCR [28] | https://pfocr.wikipathways.org | Platform for browsing pathway information extracted from published figures. | Link-out | Done |
| ProHarMeD | https://proharmed.zbh.uni-hamburg.de/ | Closing the gap between (harmonized) proteomics results and mechanotyping / drug repurposing | Plugin | Done |
| ROBUST-Web [29] | https://robust-web.net/ | ROBUST is a disease module identification tool. | Plugin | Done |
| SCANet | https://pypi.org/project/scanet/ | SCANet is an all-in-one package for single-cell profiling covering the whole differential mechanotyping workflow, from inference of trait/cell-type-specific gene co-expression modules to mechanistic drug repurposing candidate prediction. | Python package | Done |
| Seed Connector Algorithm [30] | https://github.com/bwh784/SCA | Identification of network modules by adding a minimal number of edges between the seed nodes. | Link-out | Done |
| UnPaSt | https://unpast.zbh.uni-hamburg.de | Visualizer and context explorer for unsupervised expression data bicluster results. | Plugin | Done |
| WikiPathways [17] | https://wikipathways.org | Platform for browsing and visualizing pathways. | Link-out | Done |

**Table 1.** Systems medicine tools that integrate Drugst.One listed in alphabetical order. Options are 'Link-out', referring to a URL based redirect from the tool or website to the Drugst.One standalone page, 'Plugin', referring to the integration of the javascript-based plugin into the web tool, and programmatic access using the 'Python package'.

## Drugst.One integration and customization

The Drugst.One ecosystem is a multi-component platform consisting of a website, the web plugin, a server, a content delivery system (CDS), and a Python package, as depicted in Figure 2 (for details, see Supplementary Notes 1).

The web plugin can be added to any webpage by importing one JavaScript and one stylesheet file from the https://cdn.drugst.one distribution server, and by adding the 'drugst-one' HTML tag to the source code of any system medicine tool's website (Supplementary Figure S2). Features can be customized to a high degree through JSON configuration strings that are passed as attributes. This includes default states of on/off toggles, the network, and the node and edge groups that define the network style. The plugin is responsive to changes during runtime, allowing developers to add buttons or other controls to the host page, for example switching between networks. For seamless integration of the rendered plugin into any website, styling and coloring are controllable by adding specific CSS variables to the website stylesheet. To assist developers in the integration process, the Drugst.One website provides conclusive documentation of available parameters, features, and styles. It further offers an interactive configuration page at https://drugst.one/playground where configuration options are categorized, and the replication of a configured Drugst.One instance is achieved by simple copy-pasting of the generated code snippets to the developers' websites. This low-code approach allows bioinformatics researchers to provide the community with an interactive mechanism mining web tool within hours or even minutes instead of days. The lightweight Drugst.One JavaScript library connects to the Drugst.One data warehouse server, which handles all the computationally expensive work like data annotation, mapping, and asynchronous algorithm execution.

Alternatively, a standalone integration of Drugst.One is provided at https://drugst.one/standalone, which can be accessed and customized using URLs or POST-based requests. This way, results from any website or even a command line tool can be redirected to Drugst.One through a simple web service request (Supplementary Notes 2). Detailed documentation about all Drugst.One integration options can be found at https://drugst.one/doc.

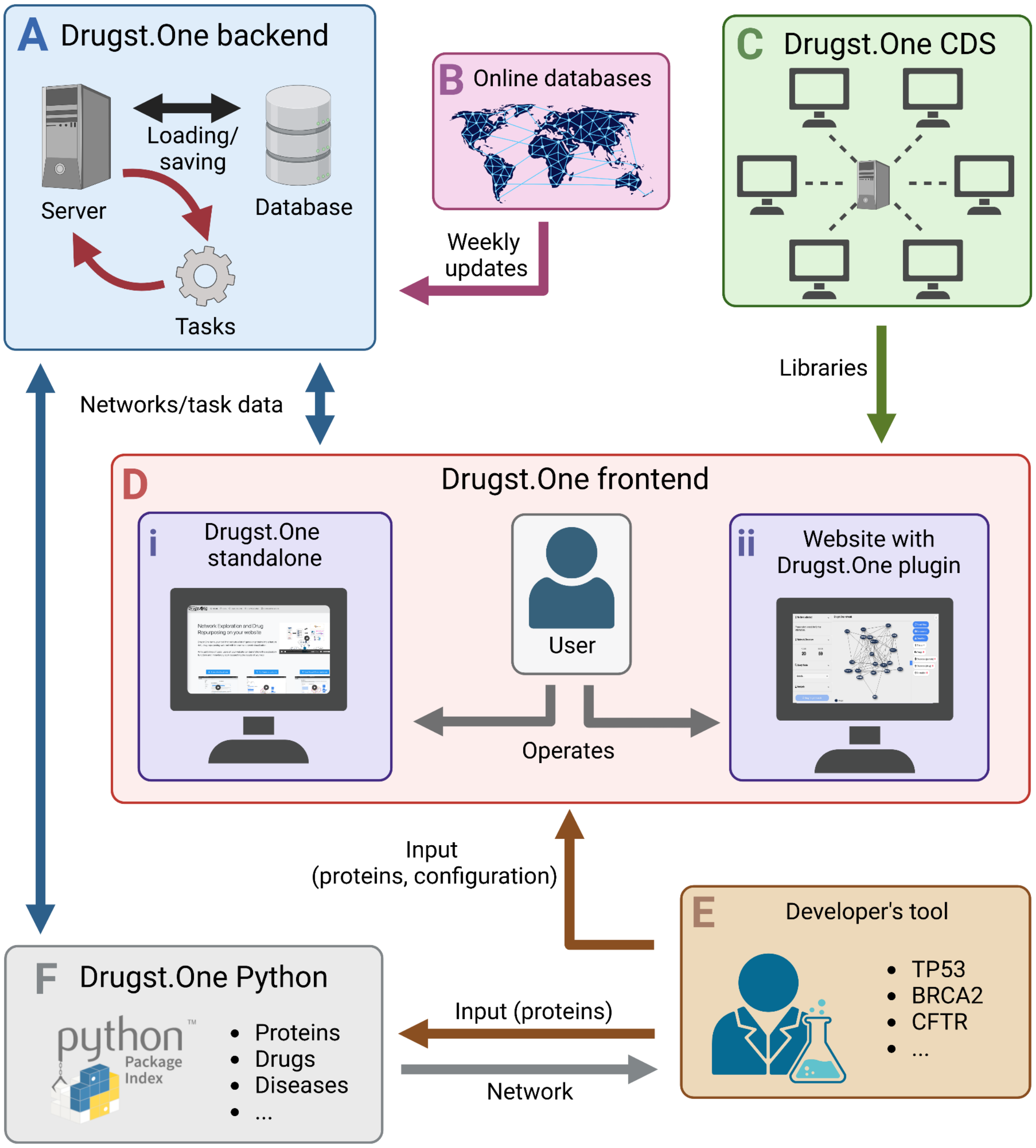

**Figure 2:** The Drugst.One ecosystem: The Drugst.One server (**A**) updates weekly from online databases (**B**), executes computationally demanding tasks, and provides data to the Drugst.One plugin (**D i** and **D ii**). The frontend is loaded from the content delivery system (CDS), (**C**), receives the network data from the developer integrating Drugst.One (**E**), and presents it to the user. Drugst.One can also be accessed programmatically through a Python package (**F**).

## Drugst.One integration examples

### Drugst.One plugin integration with ROBUST-Web

ROBUST-Web (https://robust-web.net) presents a modified version of ROBUST [31] in an online web interface. It provides a network-based disease module identification algorithm based on prize-collecting Steiner trees that mitigates study bias using edge costs derived from study-attention or bait-usage information. Given a set of seed genes and a PPI network, ROBUST-Web constructs disease modules and passes nodes and edges to the Drugst.One plugin that takes care of result presentation and visualization in an interactive network view. Drugst.One also serves as a network explorer for the analysis of modules by offering an estimation of functional coherence with DIGEST [19], GO enrichment with g:Profiler, or a lookup in NDEx IQuery for identifying pathways with the same participants. Additionally, it adds disease annotations and drug repurposing functions to make the results of ROBUST-Web more actionable and derive hypotheses for follow-up research.

### Drugst.One plugin integration with BiCoN

BiCoN [8] is a systems medicine tool for simultaneous patient stratification and disease mechanism identification, i.e., network-based endotyping. BiCoN uses a molecular interaction network as input and identifies two subgroups of patients along with a subnetwork that is enriched for differentially expressed genes between the two groups. These subnetworks can serve as composite biomarkers but may also be enriched for putative drug targets. Since BiCoN also features a web version (https://exbio.wzw.tum.de/bicon), we integrated the Drugst.One plugin for enhancing the result presentation by interactively visualizing the identified subnetworks. This allows users to explore possible drug repurposing candidates targeting the newly identified disease mechanisms, which can subsequently be experimentally validated.

### Drugst.One link-out from WikiPathways

WikiPathways [32] is a widely used, community-driven platform for exploring molecular pathways. It allows users to upload, edit, browse, and download a constantly growing pool of pathway datasets. Pathway data can be used to identify and understand key players in metabolism, which is critical for understanding rare or common diseases such as COVID-19 [32]. Thus, pathways allow for the prediction of drug target and drug repurposing candidates and are commonly used in the development of new disease treatments [33]. When inspecting individual pathways on the WikiPathways platform, users now have the option to forward the pathway genes to the Drugst.One standalone version by clicking a ‘Query Drugst.One’ link now provided by WikiPathways, located in the search menu of the ‘Participants’ table. The link redirects the user to the Drugst.One website, visualizing pathway genes, drugs directly targeting them, and offering the complete toolset of Drugst.One. In the following, we give an example of the Drugst.One usage for exploration of the smooth muscle differentiation and proliferation pathway (WP1991).

WikiPathway WP1991 describes the mechanism behind smooth muscle cell (SMC) differentiation and proliferation. The WikiPathways web interface now incorporates a button to export the pathway genes into Drugst.One (using the magnifier glass in the table showing the proteins participating in the pathway, state 03.07.23) and visualizing their interactions with drugs in Drugst.One directly. To gain a general overview of the complications (symptoms, comorbidities, etc.) associated with dysfunctional SMC development and their drivers, Drugst.One allows for extending the WikiPathways-exported network by the corresponding disorders and their associated pathway genes. Several disease nodes appear in the network, mainly representing various cardiovascular disorders (CVDs), e.g., cardiomyopathy, coronary artery disease, and aortic valve (Figure 3). The importance of SMCs for proper vascular functionality [34,35] and thus to atherosclerosis, hypertension, myocardial infarction, and other cardiovascular diseases was reported before [36–38]. An isolated subnetwork community of proteins is formed by the three myocyte enhancer factors *MEF2A*, *MEF2C*, and *MEF2D*. Besides their obvious cardiovascular implications, these factors play a role in neurological processes [39]. An impact of SMCs on status epilepticus was shown in mouse models [40], and a connection between migraine and SMC dysfunction was suggested as well [41,42].

Drugst.One allows for the projection of (gene) expression data from GTEx on the proteins in the network. The relative expression of these genes appears to be quite high in arteries and organs that have to perform physical motion, like heart, lung, bladder, and skeletal muscles, but with observable fluctuations in the relative expression of genes like *CCND2*.

With one mouse click, we import drug target information for drug repurposing candidate prediction. Despite SMCs relation to cardiovascular diseases, no corresponding CVD drugs have been identified. Mainly anti-cancer drugs (e.g., sunitinib, erlotinib, midostaurin, and ruxolitinib) targeting calcium/calmodulin-dependent protein kinase II delta (*CAMK2D)*, which is associated with cancer growth [43], are found. Notably, however, *CAMK2D* also plays a role in calcium signaling, which is essential for the upkeep of SMC function [44,45]. Hence, this may explain the observed cardiovascular side effects of *CAMK2D*-targeting drugs. According to SIDER [46], sunitinib may cause hypertensive symptoms and corresponding studies suggest that midostaurin has cardiotoxic effects [47].
Algorithms integrated in Drugst.One can extend the search space by looking for indirectly connected drugs. The selection menu offers a function to automatically add all displayed proteins to the selection, serving as the starting point (seeds) of subsequent searches. The harmonic centrality algorithm (see Supplement 3.2.3) was used to extend the network by the ten drugs with the highest score, including indirectly (transitively) connected drugs from the NeDRex database. Through this search, the tyrosine kinase inhibitor nintedanib, which has shown promising effects in pulmonary arterial smooth muscle cells and intestinal smooth muscle cells [48,49], can be identified.

This shows the identification potential of mechanism-associated drugs through the network-based drug repurposing functions Drugst.One incorporates. Whereas before only drugs primarily used in cancer were present through direct association with SMC pathway participants, Drugst.One suggested more relevant options for this context.

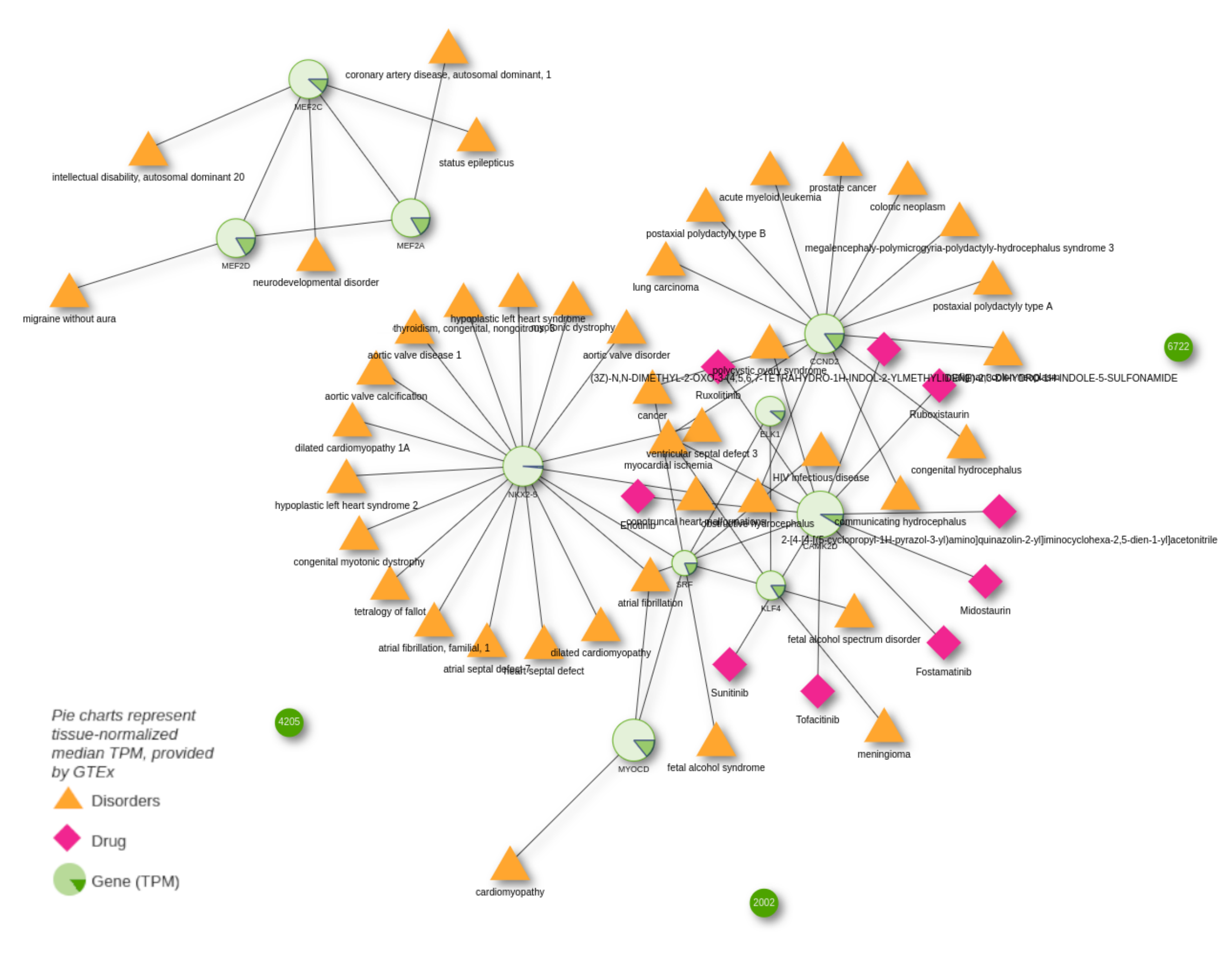

**Figure 3**: Participants of WikiPathway WP1991 displayed in Drugst.One. Adjacent diseases and drugs are enabled, as well as diseases linked to drugs targeting this smooth muscle cell proliferation and differentiation pathway. Normalized median expression values for ‘Artery - Aorta’ are overlaid as pie charts, where 360° represent the maximum observed transcripts per million (TPM) in the selected tissue and all other TPMs are exponentially scaled.

## Discussion

Biomedical research generates a wealth of data that could inform the development of novel therapies or treatments. However, despite this potential, a significant portion of the analyses conducted in this field fail to translate into clinical trials, leading to major issues in the effectiveness of public health research [50]. To this end, Drugst.One has the potential to help transform specifically omics-based research results into actionable hypotheses with potential clinical impact. Drugst.One offers a community-driven solution to streamline the knowledge distributed over many online resources for multi-omics analyses and other biomedical tools [51] to turn the results of biomedical analyses into concrete candidate drug targets and drug repurposing hypotheses. Still, we emphasize that the drug target and drug repurposing predictions are merely candidates and supervision with expert knowledge is still required before experimental validation. Drugst.One delivers explainable indications based on established biological data like expression and known disease associations or drug indications, however, the interpretation of their application in the case-specific context is up to the user. Therefore, we designed Drugst.One to be operated with maximal transparency and allow optional user input for every step of the analysis.

With the infrastructure and the resources being provided, Drugst.One helps to find a community-wide solution for standardization and streamlining the visualization of explainable disease modules and their pharmacological implications. Drugst.One provides various interfaces to be highly accessible and customizable by all members of the community while maintaining up-to-date database information and network analysis algorithms. Smooth integration into most biomedical websites and tools is confirmed by 21 resources already integrating Drugst.One. For future developers who wish to customize Drugst.One before its integration, an interactive web interface provides copy-paste-able code for customized plugin integration with their own website. An endpoint for developers who want to link out from any of their websites, apps, or command line tools is provided by Drugst.One as well.

Drugst.One complies with community standards regarding data management as defined by the FAIRness principles [52]. Download links for any data shown in Drugst.One are provided at any step, whether it is a table with drug target and drug candidates or the visualized network with all activated extensions like expression information. Export to current community standards is supported via exporting compatible .graphml files, which can be loaded directly into, e.g., Cytoscape [13]. To further increase reproducibility and interoperability, concrete plans to implement save and export functions of Drugst.One networks to NDEx [53,54] are made.
In summary, Drugst.One offers an important service to the systems medicine research community to tackle the widely recurring problem of web-based disease mechanism mining and drug repurposing candidate prediction by capturing the results of biomedical assays.

## Acknowledgments

REPO-TRIAL: This project has received funding from the European Union's Horizon 2020 research and innovation programme under grant agreement No 777111. This publication reflects only the authors' view and the European Commission is not responsible for any use that may be made of the information it contains.

RePo4EU: This project is funded by the European Union under grant agreement No. 101057619. Views and opinions expressed are however those of the author(s) only and do not necessarily reflect those of the European Union or European Health and Digital Executive Agency (HADEA). Neither the European Union nor the granting authority can be held responsible for them. This work was also partly supported by the Swiss State Secretariat for Education, Research and Innovation (SERI) under contract No. 22.00115.

This work was supported by the German Federal Ministry of Education and Research (BMBF) within the framework of "CLINSPECT-M" (grant FKZ161L0214A). This work was supported by the Technical University Munich – Institute for Advanced Study, funded by the German Excellence Initiative. This work was supported in part by the Intramural Research Programs (IRPs) of the National Institute of Diabetes and Digestive and Kidney Diseases (NIDDK). Funded by the Deutsche Forschungsgemeinschaft (DFG, German Research Foundation) – 422216132.

JB was partially funded by his VILLUM Young Investigator Grant nr.13154.

This project has received funding from the European Research Council (ERC) Consolidator Grant 770827 and the Spanish State Research Agency AEI 10.13039/501100011033 grant number PID2019-105500GB-I00.

IJ was supported in part by funding from Natural Sciences Research Council (NSERC #203475), Canada Foundation for Innovation (CFI #225404, #30865), Ontario Research Fund (RDI #34876), IBM and Ian Lawson van Toch Fund.

SL has received funding from the European Union's Horizon 2020 research and innovation programme under grant agreement No 965193 for DECIDER.

Images were created with https://www.biorender.com

## Author contributions

AM and MH developed and implemented the Drugst.One ecosystem, and wrote the manuscript. JB and OZ supervised the project and provided continuous feedback and ideas during the development. The rest of the *Drugst.One Initiative* integrated the Drugst.One ecosystem into their platforms, tested, provided helpful feedback, and revised the manuscript.
All authors agreed to the final version of the manuscript.

## Conflicts of Interest

JSR reports funding from GSK, Pfizer and Sanofi and fees from Travere Therapeutics and Astex Pharmaceuticals.

# Supplementary information

## 1. Drugst.One ecosystem

The Drugst.One ecosystem consists of the Drugst.One plugin for developers and a content delivery system (CDS) to distribute it, a website, a backend server, and a Python package (Figure 2). Any communication between Drugst.One components is SSL encrypted using HTTPS. We use GitHub for versioning of plugin and backend as well as code example repositories, all of which can be found here: https://github.com/drugst-one. For the Drugst.One website, the CDN server, CI/CD pipelines and container registries we, for now, use a GitLab instance hosted by the RRZ, the computing center of the University of Hamburg.

### 1.1 Plugin

The Drugst.One web plugin visualizes and enriches protein-protein interaction (PPI) networks (Figure S1). Given a set of gene or protein identifiers, or small networks, interaction information can be completed automatically from the Drugst.One database and the users of the website can explore the nodes in their network context. With one click, transitive connections between otherwise unconnected proteins within the interactome can be identified or first-neighbor diseases and drugs can be added to the network. Further, tissue-specific expression of the proteins can be highlighted directly in the network.

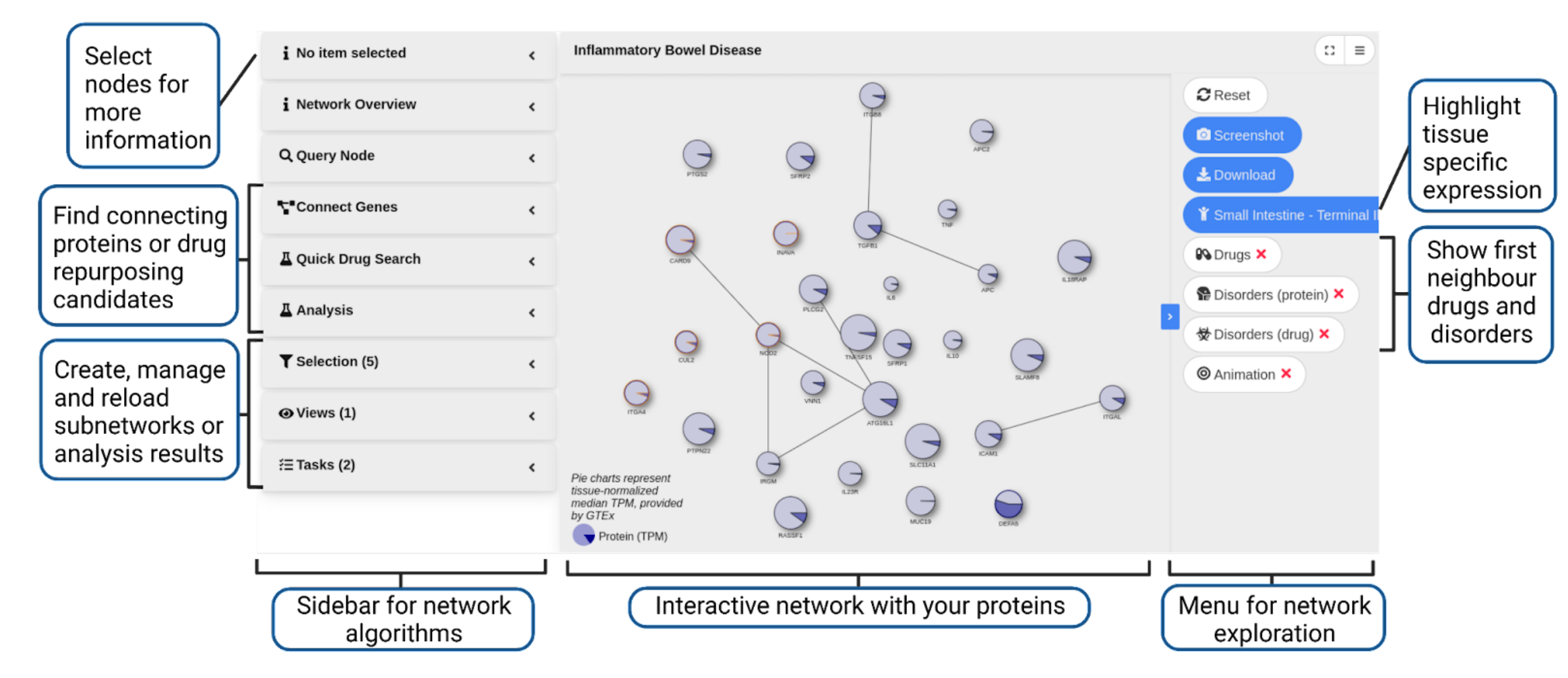

**Figure S1:** The interface of the Drugst.One plugin consists of a sidebar with network information and options for drug repurposing algorithms (left), an interactive network visualization (center), and a menu with functions to modify the network (right). Individual components and positions of these elements can be configured. Visualized are the 30 genes from the inflammatory bowel disease use case (section 4 in Supplementary Notes).

Besides exploring the loaded genes or proteins, the plugin can be used to generate drug repurposing hypotheses. Using the explorative functionalities such as highlighting proteins based on their expression in tissues, input proteins for analysis tasks can be selected. Leveraging the power of specialized network algorithms, disease modules can be identified to find additional drug targets within the same genetic context and drug repurposing candidates can be prioritized (Figure S2). Analysis results are stored for future access and can be loaded directly into the interface where a network visualization helps to understand the relation of the result to the input nodes. New analyses can be started on the original network or based on previous analysis results, supporting an iterative approach to a transparent and user-driven drug repurposing.

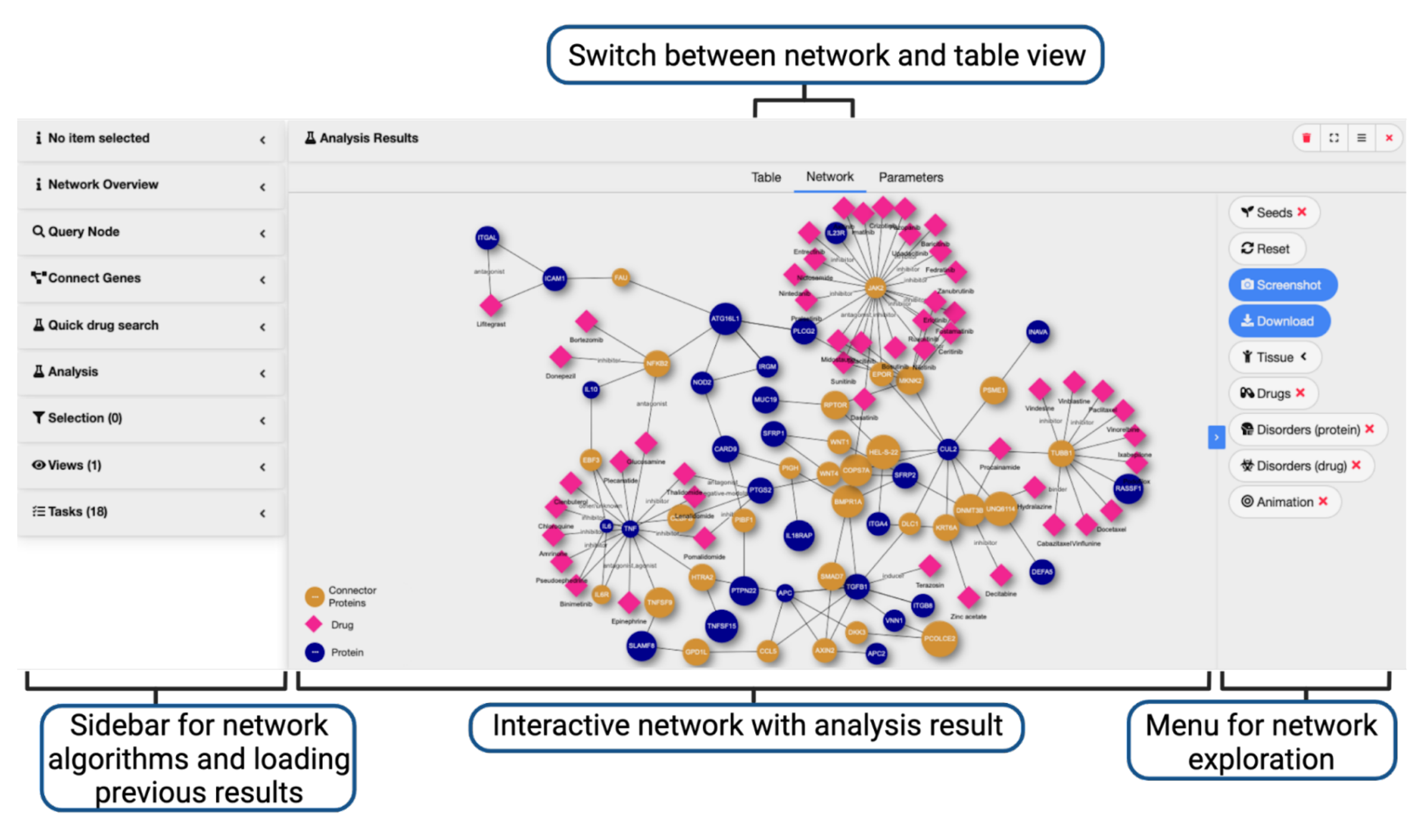

**Figure S2**: The results of drug target and drug repurposing candidates search for the inflammatory bowel disease use case (Supplements 4) are represented in the Drugst.One plugin as a network. A 'Quick Drug Search' was executed on all proteins; hence, a Multi-Steiner tree was used to connect the genes and a harmonic centrality algorithm to identify putative drug candidates.

Despite all these features the plugin remains completely serverless for hosting websites as all computations take place on the Drugst.One servers. The customizable plugin consists of a JavaScript-based web component, which can be fitted to the needs of the host website through JSON-formatted configuration strings. Functions, buttons, colors, and dimensions of the component can be adjusted to seamlessly blend in with the rest of the page. Style encapsulation guarantees no interference with styles from the host webpage. Lastly, the Drugst.One plugin is lightweight, no heavy libraries will be added to the host's webpage to minimize loading times.

The latest Drugst.One plugin (at submission v1.1.19) is developed in Angular.js 14 with TypeScript 4.4.4 and compiled and packaged with npm 8.15.0 and node 16.17.1 into

drugstone.js and drugstone.css files. For network visualization, vis.js (v9.1.6) is used and screenshot support is provided through dom-to-image-cross-origin (v.2.6.7).

## 1.2 Content delivery system

To enable the easy use and distribution of the plugin, we set up a simple content delivery system (CDS) to account for a large number of simultaneously loaded plugin instances on multiple hosting web pages. The CDS manages the plugin versions by maintaining all builds of previous plugin versions that can be accessed through fixed version identifiers. The latest stable release is tagged ‘latest’ and will be updated upon a new stable release. Unstable development versions can be tested using versions tagged as release candidates (‘-rc’). The most current development version can be accessed through the tag ‘nightly’. A list of all available plugin versions is generated after each build and is available at https://cdn.drugst.one/.

## 1.3 Website

The Drugst.One website (https://drugst.one/) serves as a documentation and exploration environment for the Drugst.One plugin. Many examples in the documentation and tutorial videos guide users through the individual features of the Drugst.One plugin as well as its integration.

The Drugst.One playground provides a graphical interface to conveniently generate custom “copy-and-paste” code snippets incorporating style and configuration choices made by the user. Further, a standalone version of the Drugst.One plugin is integrated into the website and can be accessed and modified by passing parameters using GET or POST HTTP requests, allowing developers to use the plugin without having to host a web service or incorporate the plugin in a webpage. Additionally, to facilitate and initiate the development of novel web tools presenting biomedical networks, it provides a website application template with the Drugst.One plugin.

## 1.4 Server

The Drugst.One backend server consists of a Django API (v3.2.19) connected to a PostgreSQL (v14) database in combination with a redis broker and worker (v7.0.11) and a Celery scheduler (v5.2.7). The backend interacts closely with the plugin, instantly returning information to the loaded proteins and adding the interactions between the proteins to create the PPI network. It further handles the asynchronous execution of the network algorithms in the analysis tasks by starting separate jobs on the redis server, allowing it to handle a large number of jobs in parallel or queue them accounting for the workload of Drugst.One instances on multiple websites. Most network algorithms (Supplement 3.2) are implemented with graph-tool (version 2.55), a performant C++-based network library for Python, and pre-processed network files, resulting in fast execution times of usually less than 30 seconds.

## 1.5 Python package

Programmatic access is supported using Python 3.6 and newer. The Python package (https://pypi.org/project/drugstone/) can be installed using pip (‘pip install drugstone’) and supports the main functionalities of the plugin, including fetching protein information, PPIs, and conducting drug repurposing analyses. With the Drugst.One python package, a larger number of tasks can be executed in an automated fashion, then form the GUI of the plugin, empowering developers to integrate Dugst.One into custom workflows of their own programs.

# 2. Integration

## 2.1 Plugin

Developers can add the plugin to their own website using JavaScript. This may help to visualize any results containing a list of proteins or PPIs and to add the Drugst.One functionalities. Integration is done in three steps. Firstly, the Drugst.One libraries need to be loaded with the website by integrating the import statement into the head tag of your website. It is possible to define a specific version identifier to use a static version of the plugin or to use the tag ‘latest’ to automatically update to the latest stable version upon a new release.

```
<head>
   <script src="https://cdn.drugst.one/latest/drugstone.js"></script>
   <link rel="stylesheet"
href="https://cdn.drugst.one/latest/styles.css">
</head>
```

Secondly, the Drugst.One component needs to be placed on the website. After loading the Drugst.One libraries, the html tag ‘drugst-one’ becomes available. The position of the tag defines the position of the component on a webpage.

```
<drugst-one id='drugstone-component-id'></drugst-one>
```

Lastly, the component can be configured by passing options as JSON strings to the three attributes ‘groups’, ‘config’, and ‘network’. The ‘network’ parameter can accept lists of nodes and optionally edges to construct the network. Each node and edge in the network is assigned to a node and edge group, respectively, which defines the styles of each member. Additionally, groups are assigned a group name that all its members inherit, e.g. ‘protein’ or ‘drug’. Individual nodes and edges may receive individual styles to highlight them which will override the group styles. Because the network is based on vis.js, all node and edge attributes used in vis.js are applicable, including e.g. directed edge styles. Settings regarding the component itself e.g. to add or remove certain features, or which datasets to use may be passed to the ‘config’ parameter.

```
<drugst-one
groups='{"nodeGroups":{"Protein":{"type":"Protein","color":"#ff881f","fo
nt":{"color":"#ffffff"},"groupName":"Protein","shape":"ellipse"}},
"edgeGroups":{"PPI":{"color":"#111111", "groupName":"PPI"}}}'
   config='{"identifier":"symbol","title":"Breast cancer example
network"}'
   network='{"nodes":[{"id":"BRCA1","label":"BRCA1", "group":
"Protein"}, {"id":"BRCA2","label":"BRCA2", "group": "Protein"}],
"edges": [{"from":"BRCA1", "to":"BRCA2"}]}'>
  </drugst-one>
```

The Drugst.One plugin adapts to any changes of the options immediately, allowing developers to add features like buttons, toggles, or selections to e.g. change or adjust the network at runtime.

Custom styling of the component can be achieved by setting global CSS parameters on the website. Through the prefix ‘drgstn’ it is ensured that CSS variables will not randomly collide with other style variables.

```
:root {
   --drgstn-primary:#347eee;
   --drgstn-secondary:#2e42f2;
   --drgstn-success:#48C774;
   --drgstn-warning:#ffdd00;
   --drgstn-danger:#ff2744;
   --drgstn-background:#f8f9fa;
   --drgstn-panel:#ffffff;
   --drgstn-info:#61c43d;
   --drgstn-text-primary:#151515;
   --drgstn-text-secondary:#eeeeee;
   --drgstn-border:rgba(0, 0, 0, 0.2);
   --drgstn-tooltip:rgba(74,74,74,0.9);
   --drgstn-panel-secondary:#FFFFFF;
   --drgstn-height:600px;
   --drgstn-font-family:Helvetica Neue, sans-serif;
}
```

## 2.2 Standalone

In cases where it is not desired or possible to integrate the Drugst.One component to a website, e.g., because a tool has no website, the version hosted at the Drugst.One website is accessible using HTTP requests. When working with small networks and little customization of the plugin is required, it is sufficient to encode the parameters in GET requests.

```
https://drugst.one?nodes=PTEN,TP53&edges=PTEN%20TP53&autofillEdges=false
```

For the GET-based configuration, only selected parameters are available (https://drugst.one/doc#standalone_api).
Drugst.One buttons are provided to facilitate the integration.

```html
<link rel="stylesheet"
href="https://cdn.drugst.one/libs/drugstone-buttons/0.0.1/drugstone-butt
ons.min.css">

<a class="drugstone-button drugstone-grey"
href="https://drugst.one/standalone?nodes=PTEN,TP53,BRCA2&autofillEdges=
true&activateNetworkMenuButtonAdjacentDrugs=true&interactionDrugProtein=
NeDRex&licensedDatasets=true" target="_blank">Drugst.One</a>
```

To pass more network data or extensive configuration parameters that are otherwise not available or exceed the URL limit of 2048 characters, POST requests can be used. An API endpoint expects the same options as the plugin and returns an identifier with which the network can be loaded using a GET request (i.e. an URL):

```
(Send options to API and GET identifier)
let networkID = post(
   'https://api.drugst.one/create_network',
   {
       network: {nodes: [...], edges: [...]},
       groups: {...}
       config: {...}
   }
)
```

```
(Load configuration in Drugst.One standalone with a network identifier)
https://drugst.one?id=<networkID>
```

# 3. Methods

## 3.1 Data integration

An essential contribution of Drugst.One is the integration of multiple data sources that can be selected to add information to visualized data. Basic entities that are considered in Drugst.One are proteins/genes, drugs, and diseases. The available ID spaces for gene or protein entities are HGNC [1], UniProt [2], Ensemble [3], and Entrez [4]. For drugs Drugst.One uses DrugBank and for disorders MONDO [5] identifiers. To describe links between the different entities Drugst.One integrates four different relational layer types, namely protein-protein, protein-drug, protein-disorder, and drug-disorder data, derived from multiple different data sources (Supplementary Table 1). Another distinction between static and updating datasets can be made. Using the secondary database NeDRexDB [6], which is updated on a weekly basis, any data imported from it is automatically updated weekly using celery-beat as a scheduler. The NeDRex datasets for protein-protein and drug-target interaction data represent a combination of all individual data sources. Data that is not available in NeDRex do not receive regular updates. Some data sources have restrictive reuse licenses attached, e.g., for use in a commercial scenario. In Drugst.One, we provide both, licensed and openly available datasets, but the access to licensed data has to be unlocked with a configuration parameter. At the time of publication, the following datasets and their respective the end-user license agreements (EULA) are available in the Drugst.One plugin menu:

| Source | Version | Layers | Licensed |
|---|---|---|---|
| APID [7] | January 2019 | Protein-Protein | no |
| BioGRID* [8] | 2023-07-03 | Protein-Protein | no |
| ChEMBL [9] | 27 | Protein-Drug | no |
| CTD* [10] | 2023-07-03 | Drug-Disorder | no |
| DGIdb [11] | 4.2.0 | Protein-Drug | no |
| DisGeNET*[12] | 2023-07-03 | Protein-Disorder | no |
| DrugBank [13] | 5.1.8 | Drug-Disorder | yes |
| DrugBank* [13] | 2023-07-02 | Protein-Drug | yes |
| DrugCentral* [14] | 2023-07-03 | Protein-Drug, Drug-Disorder | no |
| GTEx [15] | v8 | Tissue Expression | no |
| IID* [16] | 2023-07-03 | Protein-Protein | no |
| IntAct* [17] | 2023-07-03 | Protein-Protein | no |
| NeDRex [6] | 2.10.0 | Protein-Protein, Protein-Drug, Protein-Disorder, Drug-Disorder | yes |
| NeDRex [6] | 2.10.0 | Protein-Protein, Protein-Drug, Protein-Disorder, Drug-Disorder | no |
| OMIM* [18] | 27-12-2022 | Protein-Disorder | yes |
| STRING [19] | 11.0 | Protein-Protein | no |

**Supplementary Table 1.** List of association types and source databases stored in the Drugst.One data warehouse. The version refers to the latest state before submission. *Databases with an asterisk are integrated as a part of NeDRexDB.

## 3.2 Algorithms

Seven network mining algorithms are implemented in Drugst.One for module identification and/or drug prioritization. To keep the plugin lightweight and easy to use, algorithms were deliberately chosen due to their focus on basic network properties like degree centrality and network proximity. Depending on the type of network, e.g. sparsely or densely connected, users can try out different approaches to explore the search space of related drug targets and drug repurposing candidates. While all of the algorithms share some general parameters (Supplement 3.2.1), some of the algorithms may offer additional settings (Supplement 3.2.2 - 3.2.9).

### 3.2.1 General parameters

For each algorithm the following options exist:

- Result Size. The number of returned nodes (drug targets or drugs)
- Maximum Degree. Option to filter out hub genes. If set to > 0, genes with a network degree in the complete gene-drug interaction network above this threshold will be excluded.
- Hub Penalty. Penalizes genes with a large degree in the network.
- Filter Edges. If set, only the shortest paths to the drug target or drug nodes will be displayed in the resulting network. Otherwise, all found pathways will be shown.

### 3.2.2 Betweenness centrality

Betweenness is obtained by finding the shortest paths for each pair of nodes in the network and assessing the number of shortest paths that pass through a particular node, such that a measure of the centrality of a node in a network global context is received. Betweenness centrality has been established as a common measurement in network biological applications [20] and is especially practical in finding communities in large networks [21]. In Drugst.One, betweenness is based on the shortest paths between the seed nodes only and can be used to find drug targets with maximized connectivity to all seeds.

### 3.2.3 Harmonic centrality

Harmonic centrality ($C_h$) measurement can be described as the average shortest distance from each node to all other nodes in a network. This measurement is the equivalent of harmonic centrality for disconnected graphs. Formally speaking, it can be annotated as

$$C_h = \sum_{y \neq x}^{y} \left( \frac{1}{dist(x,y)} \right)$$

where $x$ is a given node and $\frac{1}{dist(x,y)} = 0$ if $dist(x,y) = \infty$ [22]. The closer a node is to other nodes, the higher the score. It has already been proven successful in a number of biological network problems for instance with metabolic or PPI networks [23,24].

### 3.2.4 Degree centrality

Degree centrality ($C_d$) measurement is obtained by ranking the nodes in a network based on their degree, which is defined as the number of neighbors a node has divided by the total number of nodes in the network. It can be described as

$$C_d(x) = deg(x)$$

where x is a given node and $deg(x)$ is its degree. While it is a commonly used network analysis technique, it most importantly has been shown useful in the identification of essential proteins in PPI networks [25,26]. Thus, it is a simple approach for classifying the network-related importance of a particular protein. In Drugst.One, it can be used to discover valuable drug targets or drugs, based on the seed selection given by the user. However, ranking by node degree is prone to introduce research bias.

### 3.2.5 KeyPathwayMiner

KeyPathwayMiner (KPM) is an online tool developed by Alcaraz et al. for pathway enrichment analysis [27]. Users can utilize KPM for their drug target search by selecting seed genes from the network and letting KPM find an interaction network of genes spanned by the seed genes. The resulting genes are functionally related to the seed nodes and therefore are suitable drug target candidates. Only one parameter $k$ has to be set by the user, which defines the amount of permitted intermediate nodes that are neither part of the seed nodes nor the common pathway.

- Additional proteins $k$: Number of intermediate nodes allowed between the seed nodes.

### 3.2.6 Multi-Steiner tree

The Multi-Steiner tree algorithm [28] approximates the minimum spanning tree connecting the seed nodes in a reasonable time. The implementation is adopted from Ahmed et al. [29]. It can be used to create a minimum spanning subnetwork between user-selected seed nodes, which happen to be central interaction partners between the seed nodes and thus represent favorable drug targets. The user can specify the number of Steiner trees computed to approximate a minimum spanning tree, and the tolerance indicating how much the subsequent trees may increase the number of edges a higher number of Steiner trees leads to more variations at the cost of a longer runtime.

### 3.2.7 Network proximity

As introduced by Guney et al. [29], network proximity is the average length of shortest paths from drug target nodes to all of the user-selected seed nodes. The algorithm then computes a statistical significance score compared against random expectation. This algorithm was adopted in Drugst.One so that best-scored drugs are returned to the user as candidate drugs.

### 3.2.8 TrustRank

TrustRank [30] is based on the same concepts as the Google PageRank algorithm and harmonic centrality [31,32]. A trust score is propagated through the network starting at the seeds, damping the score based on the distance traveled. The user can set the damping factor in a range from 0-1, with a higher damping factor causing the propagation to either stop at nodes in close proximity or in larger portions of the network. In Drugst.One, TrustRank is used to identify putative drug targets as well as drug candidates.

- Damping Factor: Correlates with the distance a trust score propagates through the network. The larger the factor, the larger the proportion of the network that is considered.

### 3.2.9 Algorithm applications

| **Name** | **Drug target search** | **Drug search** |
|---|---|---|
| Betweenness centrality | yes | no |
| Harmonic centrality | yes | yes |
| Degree centrality | yes | yes |
| KeyPathwayMiner | yes | no |
| Multi-Steiner tree | yes | no |
| Network proximity | no | yes |
| TrustRank | yes | yes |

Supplementary Table S2. An overview of all integrated algorithms regarding their availability in drug target and drug search functions.

# 4. Use Case: IBD

## 4.1 Repurposing of JAK inhibitors against IBD

A central aspect of Drugst.One is the focus on *in-silico* drug repurposing candidate prediction. To this end, we replicate an exemplary repurposing case study for inflammatory bowel disease (IBD).

Sadegh et al. [6] identified fostamatinib, ruxolitinib, and imatinib for application in IBD by starting from 30 seed genes associated with IBD according to DisGeNET [12] and OMIM [18] (Supplementary Figure S1, Table S3). Multi-Steiner tree (MuST) [28] was applied to connect the seed genes in the network and the closeness centrality (CC) algorithm was employed to identify drug repurposing candidates (Supplementary Figure S2).

To reproduce this example case in Drugst.One (see Table S3), we loaded the 30 seed genes associated with IBD into Drugst.One. As protein-protein and drug-target interaction datasets the licensed NeDRex versions were used and we executed a 'Quick Drug Search', consisting of a 'Drug target search' using MuST (trees=5, tolerance=5, hub penalty=0.5) to connect all seeds. Drugs were ranked using harmonic centrality, being the closest to CC used in the original paper, and the top 50 drugs were chosen as the most promising candidates. In the referenced paper, the authors identified fostamatinib, ruxolitinib, and imatinib on ranks 1, 5, and 12 respectively as drugs that have literature support for being relevant for IBD. With Drugst.One, the same drugs are found at ranks 4, 4, and 9, (Supplementary Tabe S5). The difference might stem from small variations in the interactome and drug-target data or the use of harmonic centrality instead of closeness centrality. The drugs target *JAK2*, a gene added by MuST, with ruxolitinib and fostamatinib being known JAK inhibitors (JAKi), and their potential for IBD treatment is currently under investigation [33]. Further, they inhibit *MKNK2*, another gene identified by MuST and investigated for its role in different types of colitis [34]. JAKis interact with *MAPK*, a gene well-known for its role in IBD [35,36], with which the observed effect of a dysregulated *MKNK2* might be explained. Tofacitinib, another JAKi and ranked second by HC, has the same targets as fostamatinib and ruxolitinib and is subject to studies investigating beneficial effects in IBD [37,38].

Dasatinib, the tyrosine kinase inhibitor on rank one, is known to induce ulcerative colitis, part of the IBD umbrella, in some patients [39,40]. Even though this drug does not have the desired effect, it is directly associated with the targeted disease and its pathways and may lead to the identification of impactful targets.

Rank two is shared by three immunomodulatory drugs thalidomide, pomalidomide and lenalidomide, as well as glucosamine. Thalidomide has known indications for ulcerative colitis, a subtype of IBD, while also lenalidomide and pomalidomide showed protective effects against IBD in mouse models [41]. Lopez-Millan et al. identified lenalidomide as the more potent option, suggesting its usage as a therapeutic drug against inflammatory diseases. There is evidence of beneficial effects of these immunomodulators in human IBD, but their use in clinical practice is under discussion due to severe adverse side effects [42,43].

Binimetinib is ranked third, which is mainly used as an anti-cancer drug. In this application, inflammatory colitis was observed as an adverse side effect [44], hinting towards a cause-effect relationship that can be studied further to research the mechanistic origins of

IBD.
On rank four, together with ruxolitinib and fostamatinib, other inhibitors with the same targets (*MKNK2* and *JAK2*) are found, namely tofacitinib (FDA-approved for ulcerative colitis), sunitinib, midostaurin, erlotinib, and ceritinib.
On ranks two to five, tumor necrosis factor-$\alpha$ (*TNF*) inhibitors are found. *TNF* is a known target to treat IBD [45,46] and thus effects of the drugs chloroquine on ulcerative colitis [47] and plecanatide on colitis symptoms in mice [48,49]. Further, epinephrine, pseudoephedrine, and clenbuterol share rank five, all have *TNF*-inhibiting effects [50–52] and appear to have general anti-inflammatory effects [53], most likely induced by downregulated or inhibited *IL-6* expression [52,54].
Rank six terazosin, a drug inhibiting *TGFB1*, whose dysregulation is closely linked to IBD [55].
In summary, with Drugst.One we were not only able to re-identify three promising repurposing candidates from a previous study for IBD, but show that the first 5 ranks (that includes 20 drugs) contain valid candidates or already approved drugs for IBD treatment, including hypotheses about their molecular relationship with IBD.

The following 30 seeds are IBD-associated genes that were used by Sadegh et al. [6] to build their use case.

| | | | | |
|---|---|---|---|---|
| ATG16L1 | ICAM1 | TNF | SFRP2 | APC2 |
| IL10 | CUL2 | SFRP1 | TNFSF15 | ITGA4 |
| DEFA5 | MUC19 | SLAMF8 | APC | IL6 |
| INAVA | CARD9 | ITGB8 | IL23R | NOD2 |
| RASSF1 | SLC11A1 | IL18RAP | TGFB1 | IRGM |
| PLCG2 | PTPN22 | PTGS2 | VNN1 | ITGAL |

Supplementary Table S3. List of the 30 IBD-associated genes used by Sadegh et al. [6] in their drug repurposing case study. This list serves as input for an example use case highlighting the potential of Drugst.One.
The parameters that can be used to replicate the use case, may be found in table S4.

| Drugst.One | |
|---|---|
| Protein-Protein interaction dataset | NeDRex (licensed) |
| Drug-Protein (target) interaction dataset | NeDRex (licensed) |
| **Drug target search** | |
| Algorithm | MuST |
| Number of Steiner trees | 5 (default) |
| Tolerance for trees | 5 |
| Hub penalty | 0.5 |
| **Drug search** | |
| Algorithm | Harmonic centrality |
| Result size | 50 |

Supplementary Table S4. The Drug-Protein, as well as the Protein-Protein datasets, used for the IBD drug repurposing use case, were set to the most complete one (NeDRex - licensed). For reproducibility, the exact parameters used in drug target and drug identification steps are listed.

| Drug | Score | Rank |
|---|---|---|
| Dasatinib | 1 | 1 |
| Lenalidomide | 0.9882636549013167 | 2 |
| Thalidomide | 0.9882636549013167 | 2 |
| Glucosamine | 0.9882636549013167 | 2 |
| Pomalidomide | 0.9882636549013167 | 2 |
| Binimetinib | 0.9824981860163579 | 3 |
| Erlotinib | 0.9600936780691096 | 4 |
| Sunitinib | 0.9600936780691096 | 4 |
| Ruxolitinib | 0.9600936780691096 | 4 |
| Tofacitinib | 0.9600936780691096 | 4 |
| Midostaurin | 0.9600936780691096 | 4 |
| Fostamatinib | 0.9600936780691096 | 4 |
| Ceritinib | 0.9600936780691096 | 4 |
| Chloroquine | 0.9600936780691095 | 5 |
| Epinephrine | 0.9600936780691095 | 5 |
| Clenbuterol | 0.9600936780691095 | 5 |
| Pseudoephedrine | 0.9600936780691095 | 5 |
| Amrinone | 0.9600936780691095 | 5 |
| Plecanatide | 0.9600936780691095 | 5 |
| Terazosin | 0.9492702822172273 | 6 |
| Bortezomib | 0.9334851363762997 | 7 |
| Donepezil | 0.9334851363762997 | 7 |
| Lifitegrast | 0.9283394387790588 | 8 |
| Pralsetinib | 0.9232501599628847 | 9 |
| Nilotinib | 0.9232501599628847 | 9 |
| Niclosamide | 0.9232501599628847 | 9 |
| Fedratinib | 0.9232501599628847 | 9 |
| Imatinib | 0.9232501599628847 | 9 |
| Bosutinib | 0.9232501599628847 | 9 |
| Crizotinib | 0.9232501599628847 | 9 |
| Nintedanib | 0.9232501599628847 | 9 |
| Upadacitinib | 0.9232501599628847 | 9 |
| Entrectinib | 0.9232501599628847 | 9 |
| Pazopanib | 0.9232501599628847 | 9 |
| Zanubrutinib | 0.9232501599628847 | 9 |
| Axitinib | 0.9232501599628847 | 9 |
| Baricitinib | 0.9232501599628847 | 9 |
| Procainamide | 0.9083117074801125 | 10 |
| Decitabine | 0.9083117074801125 | 10 |
| Hydralazine | 0.9083117074801125 | 10 |
| Zinc acetate | 0.9083117074801123 | 11 |
| Vinflunine | 0.8986184380435127 | 12 |
| Ixabepilone | 0.8986184380435127 | 12 |
| Vinblastine | 0.8986184380435127 | 12 |
| Podofilox | 0.8986184380435127 | 12 |
| Paclitaxel | 0.8986184380435127 | 12 |
| Docetaxel | 0.8986184380435127 | 12 |
| Cabazitaxel | 0.8986184380435127 | 12 |
| Vindesine | 0.8986184380435127 | 12 |
| Vinorelbine | 0.8986184380435127 | 12 |

Supplementary Table S5. The top 50 drugs resulting from the 'Quick Drug Search' in the IBD drug repurposing use case. The search was conducted on the NeDRex dataset (licensed). Listed are the drugs with their respective score (normalized) as returned by algorithm as well as the assigned rank.

## 4.2 Drug candidate identification and mechanism mining through microRNA targets

The database of human microRNA (miRNA) target predictions, mirDIP (version 5.3.0.1, database version 5.2.3.1) [56], is a resource for miRNA-based regulation information. mirDIP allows the identification of gene regulation through miRNAs while avoiding a prediction bias. Its unidirectional search function can be used to either find miRNAs targeting a given set of genes or to retrieve the set of targeted genes given a set of miRNAs. In both cases, mirDIP now offers the option to visualize and explore the used or found genes using the Drugst.One plugin.

We took 30 IBD-associated genes from Sadegh et al. [6] (Supplementary Table S3) as mirDIP input to identify all known or predicted miRNA regulators ('miRNA-gene matrix' -> 'Search gene symbols'). Three microRNAs (*hsa-miR-142-3p*, *hsa-miR-3942-5p*, and *hsa-miR-574-3p*) were deemed to be main regulators of IBD genes because they each target three IBD-associated genes. Interestingly, *hsa-miR-142-3p* levels have been found to be elevated in the saliva of ulcerative colitis (UC) patients [57]. For these microRNAs, we identified the targets using mirDIP unidirectional search and loaded the list of all targeted genes into the Drugst.One plugin to identify relevant drugs and disorders (Figure S3).

Using the first neighbor drug and first neighbor drug-disease association annotation in Drugst.One a number of drugs can be found, that have indications for UC but are not targeting any of the 30 IBD-associated genes but other genes that are regulated by at least one of the three miRNAs. The drugs are sulfadiazine, azathioprine, methylprednisolone, cortisone acetate, budesonide, prednisone/prednisolone, sulfasalazine, loperamide, and hydrocortisone. These drugs target seven genes, namely *NR3C1*, *CALM1*, *RAC1*, *SLC7A11*, *HTR2A*, *SCN3A*, and *GRIN2A*.
Comorbidity between UC, or any other IBD disorder, and diseases associated with those seven genes would give an indication for a shared underlying mechanism that can be dysregulated by the investigated microRNAs. Comorbidities could not be found, even though evidence for their participation in UC, Crohn's disease, or general IBD exists (*NR3C1* [58], *RAC1* [59], *SLC7A11* [60], *HTR2A* [61]). Some indications are given by carvedilol, a hypertensive drug. It has (pre-)clinical implications for IBD and targets both *HTR2A* and *SLC7A11* [62,63].
This leaves room for further investigation of the mechanistic role of identified genes, especially *SLC7A11* and *HTR2A*, in IBD.

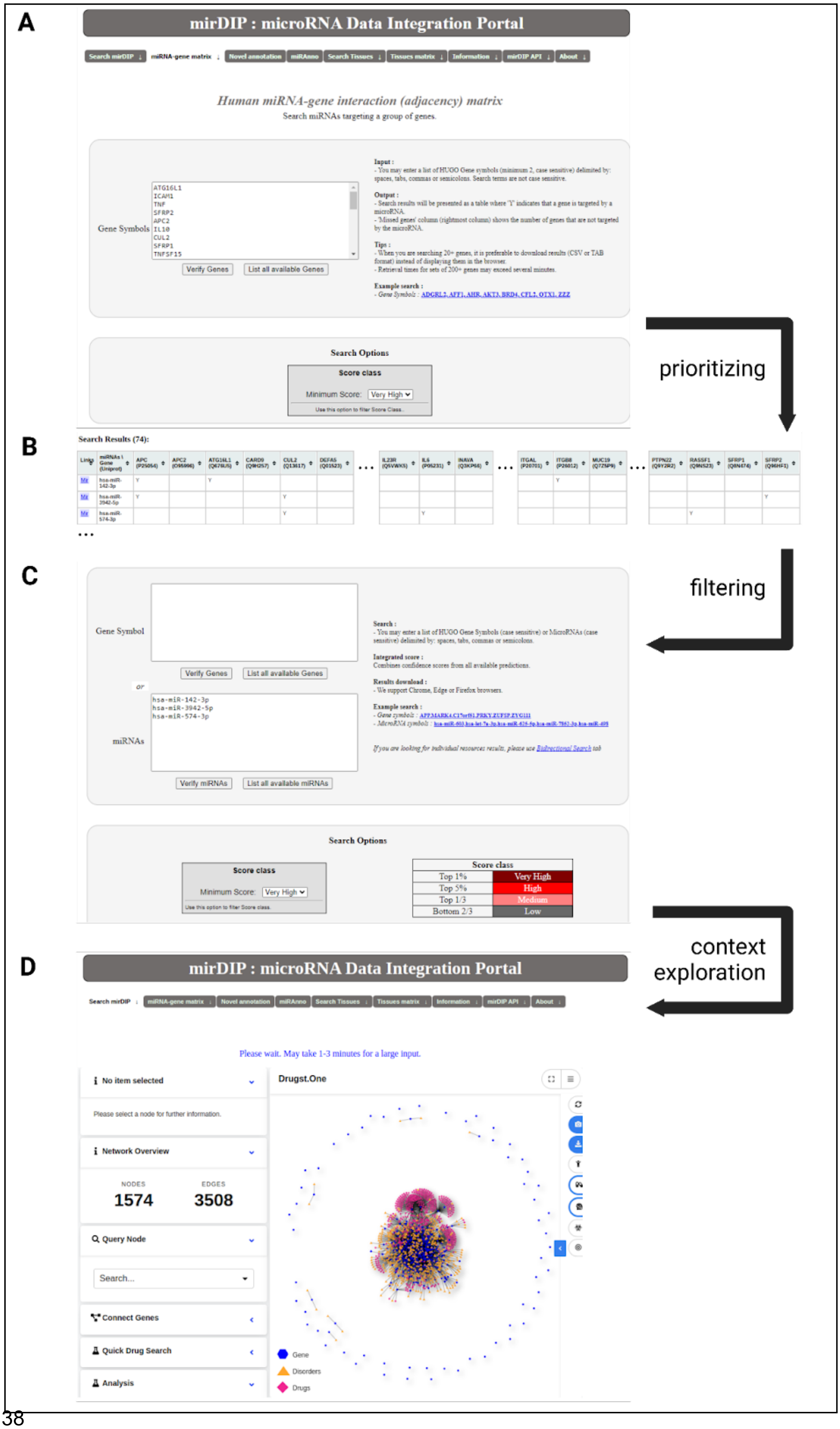
A
mirDIP : microRNA Data Integration Portal
Search mirDIP
miRNA-gene matrix
Novel annotation
miRAnno
Search Tissues
Tissues matrix
Information
mirDIP API
About
Human miRNA-gene interaction (adjacency) matrix
Search miRNAs targeting a group of genes.
Gene Symbols
ATG16L1
ICAM1
TNF
SFRP2
APC2
IL10
CUL2
SFRP1
TNFSF15
Verify Genes
List all available Genes
Input :
Output :
Tips :
Example search :
Search Options
Score class
Minimum Score:
Very High
prioritizing
B
Search Results (74):
...
C
filtering
Gene Symbol
Verify Genes
List all available Genes
or
hsa-miR-142-3p
hsa-miR-3942-5p
hsa-miR-574-3p
miRNAs
Verify miRNAs
List all available miRNAs
Search :
Integrated score :
Combines confidence scores from all available predictions.
Results download :
Example search :
Search Options
Score class
Minimum Score:
Very High
Score class
Top 1%
Very High
Top 5%
High
Top 1/3
Medium
Bottom 2/3
Low
D
context exploration
mirDIP : microRNA Data Integration Portal
Please wait. May take 1-3 minutes for a large input.
No item selected
Please select a node for further information.
Network Overview
NODES
1574
EDGES
3508
Query Node
Search...
Connect Genes
Quick Drug Search
Analysis
Drugst.One
Gene
Disorders
Drugs

**Figure S3:** Workflow using mirDIP portal with Drugst.One plugin to identify microRNA targets for the 30 IBD genes (Table S3), and related drug targets. mirDIP was queried to identify regulating miRNAs, using all databases (**A**). 74 microRNAs were identified, of which only three regulate three IBD-associated genes each (**B**). Using the three microRNAs as a query to mirDIP (**C**) identifies 534 target genes. This set of genes is used to interrogate Drugst.One (**D**).

## 5. Other Collaborations

The Drugst.One initiative has collaborations with other projects to further extend its capabilities.

| Tool | URL | Tool Description | Collaboration |
|---|---|---|---|
| BioCypher [64] | https://biocypher.org | Drugst.One uses BioCypher to facilitate the extensibility of the database and offer additional datasets, e.g. Omnipath | Integration in Drugst.One data warehouse |
| NDEx IQuery [65] | https://www.ndexbio.org/iquery/ | Web tool for pathway and network-based gene set analysis. Allows Drugst.One users to search for curated pathways based on selected genes | Integration in Drugst.One plugin |
| NDEx [66,67] | https://www.ndexbio.org/ | Web platform for storing, sharing, and publishing user-created biological networks. | Integration of “Export to NDEx” function into Drugst.One plugin |

Supplementary Table S6. Additional already initiated collaborations that are upcoming improvements to the Drugst.One platform.